\documentclass[review,number,sort&compress]{elsarticle}
\usepackage{graphicx}
\usepackage{sectsty}
\usepackage{amsmath}
\usepackage{mathtools}
\usepackage[font=footnotesize]{caption}
\usepackage{float}

\usepackage{lineno}
\def\-{\scalebox{0.9}[1.0]{-}}
\usepackage{footnote}

\begin{document}

%\linenumbers
%\title{Particle-level simulation of radio scattering from particle shower plasmas in GEANT4}
\title{Particle-level model for radar based detection of high-energy neutrino cascades}
\author[ku]{S~Prohira\corref{cor1}}%\thanks{'Corresponding author, now at The Ohio State University, prohira.1@osu.edu'}
\author[ku,mephi]{D~Besson}
\address[ku]{U. of Kansas, Lawrence, KS, U.S.A.}
\address[mephi]{National Research Nuclear University, Moscow Engineering Physics Institute, 31 Kashirskoye Highway, Russia 115409}
%\author{author list?}
\cortext[cor1]{Corresponding author, now at The Ohio State University, prohira.1@osu.edu}
%\affil {\small U. of Kansas, Lawrence, KS, U.S.A.}
%\date{}

%\bibliographystyle{unsrt}

%\ead{s742p674@ku.edu}
\begin{abstract}

  We present a particle-level model for calculating the radio scatter of incident RF radiation from the plasma formed in the wake of a particle shower.
  We incorporate this model into a software module (``RadioScatter''), which calculates the collective scattered signal using the individual particle equations of motion, accounting for plasma effects, transmitter and receiver geometries, refraction at boundaries, and antenna gain patterns. We find appreciable collective scattering amplitudes with coherent phase for a range of geometries, with high geometric and volumetric acceptance. Details of the calculation are discussed, as well as the implementation of RadioScatter into GEANT4. A laboratory test of our model, currently scheduled at SLAC in 2018, with the goal of measuring the time-dependent characteristics of the reflecting plasma, is also described. Prospects for a future in-ice, high-energy neutrino detector, along with comparison to current detection strategies, are presented. 
%\cite{t486}
%\cite{askaryan}
%\cite{gorham}
%\cite{raizer}
%\cfoot{test}

\textit{Keywords:} neutrino, particle-shower, radio-frequency, GEANT4, radio
\end{abstract}
\maketitle
%\begin{multicols}{2}
\part*{Part One: Particle-level model; RadioScatter; lab tests }
\section{Introduction} 
\setlength{\abovedisplayskip}{1cm}
\setlength{\belowdisplayskip}{1cm}
High-energy particles incident on dense media will produce a shower of secondary particles. As these shower particles traverse the interaction medium, they eject cold ionization electrons from atoms in the bulk, forming a tenuous particle-shower plasma (PSP), distinct from the energetic shower front particles responsible for ionization. For high incident particle energies ($E\geq$ 1~PeV, consistent with, and beyond, the experimental reach of the IceCube\cite{icecube} experiment), this plasma will become dense enough to reflect incident radio-frequency (RF) radiation\cite{blacket}. It has been recently suggested\cite{gorham}\cite{chiba}\cite{krijnkaelthomas}\cite{krijn_radar_18} that this technique could be used to advantage in the field of high-energy neutrino physics, where low fluxes and small interaction cross-sections demand large detection volumes. In the radio scatter approach, a large volume of interaction medium, such as ice, is illuminated with radio-frequency (RF) energy by a transmitter (TX), and any PSP of sufficient density within this volume will reflect the incident RF to a distant receiver (RX). Several experimental tests have been made to detect this phenomenon\cite{chiba2}%\cite{belz_els}
\cite{krijn_els}, but none have approached the incident particle energies, and therefore densities, of a true high-energy neutrino/ice interaction. It is this scenario that we discuss here.

There are several advantages of the radio scatter method over the current RF-based detectors for high-energy neutrinos, including ARA\cite{ara}, ARIANNA\cite{arianna} and ANITA\cite{anita}. Those experiments seek to detect primary ``Askaryan''\cite{askaryan_orig} emissions from the showers themselves. ``Askaryan radiation'' \cite{askaryan}\cite{t486} denotes collective Cherenkov radiation, confined to a cone of angular thickness $\sim$1 degree, beamed at the usual Cherenkov angle. Detection of such emission is therefore constrained to the limited solid angle of the cone, significantly limiting the geometric aperture. The radio scatter method does not suffer from this geometric limitation, and has acceptance over a much larger portion of the solid angle surrounding a high-energy neutrino shower axis. Additionally, whereas Askaryan signals are directly proportional to the energy of the primary neutrino, the radio scatter signal scales with both the neutrino energy as well as the output power of the sounding transmitter, such that a strong transmitter can effectively lower the neutrino energy threshold. The impulsive signal shape of Askaryan emission is also easily mimicked by anthropogenic transients, particularly at the South Polar ARA site which is in close proximity to the Amundsen-Scott South Pole station, making background rejection challenging. The return signal from the radio scatter method would be a characteristic, coherent, $\cal{O}$(10~ns) burst of RF with frequency content set by the transmitter-shower-receiver geometry, permitting a well-tailored firmware trigger. 

The PSP itself is a unique physical system. % (Figure~\ref{psp}). 
The cold ionization electrons are quasi-stationary, with energies of ${\cal O}$(10~eV) and an electron number density $n_e$ decreasing longitudinally at a rate set by the ionization electron lifetime, while the shower front which produces them advances at $\beta\sim~1$. The lifetime of the PSP electrons (called the plasma lifetime $\tau$) is medium-specific, and has not been experimentally verified. The best existing measurement of the ionization lifetime in ice is given in \cite{ice_properties}, and is $\cal{O}$(1-10~ns), with the lifetime dependent on the temperature and purity of the ice.
Note that $\tau$ refers to the average time required for individual free PSP electrons to be captured by positive ions or neutrals in the medium, in contrast to the much-longer lifetime of the shower itself.
For our proposed in-ice experiment, the lifetime of the plasma electrons is not well-established; nevertheless,
(as detailed below) our simulations indicate detectable, coherent radar returns
for PSP lifetimes as short as 0.1 ns. Laterally, 90\% of the shower particles are contained within 1 Moli\`ere radius from the shower axis, which for ice is order 10~cm.
%amy comment-.1 ns is from physics or what? physics wise it is probably smaller, but from an experimental standpoint, .1ns is resolvable by a 10gs/s scope, for example. 

Direct radio (Askaryan) emission from acceleration of the shower particles themselves is currently neglected in the RadioScatter module.
This is due to the fact that it will be largely beamed within a few degrees of the Cherenkov angle, and therefore only comprises a small percentage of the detectable solid angle. %Additionally, the number of ionization electrons exceeds that of shower particles by several orders of magnitude.
Reflections from the relativistically moving shower particles are also neglected, as these predominantly manifest at frequencies beyond the range of our planned data acquisition system (DAQ), and are several orders of magnitude lower in number than the ionization electrons. % through inverse Compton scattering.
%due to inverse Compton scattering, the scattering frequencies are well outside the sampling frequency range, and therefore unobservable for any reasonable sampling rate. 

% \begin{figure}[H]
% \begin{centering}
% \includegraphics[width=.7\textwidth]{psp}
% \par\end{centering}
% \caption{Graphical representation of a particle shower plasma (PSP). A stationary, short--lived PSP with a strong density gradient trails a relativistically advancing shower front.}
% \label{psp}
% \end{figure}

%other treatments use distribution functions for velocities, and assume that the plasma is stable over longer timescales
Several macroscopic models for radio scattering, treating the PSP monolithically, have been presented elsewhere\cite{gorham}\cite{krijnkaelthomas}\cite{krijn_radar_18}\cite{stasielak}\cite{bakunov_disc}. Although computationally economical, such models require assumptions regarding the development and characteristics of the plasma. %Other treatments use distribution functions for velocities, and assume that the plasma is stable over longer timescales. %Furthermore, well-established and experimentally verified models \cite{plasma_diagnostics} for radio scattering from quasi-static plasmas assume that the plasma is stable over long timescales or length scales, which cannot be assumed in the PSP case. 
Here, we calculate the reflected radar signal from the PSP microscopically, by summing over the individual scatterers in showers produced by Monte-Carlo simulations such as GEANT4\cite{geant}, accounting for charge motion in a plasma using the single-particle equation of motion (EOM).
Particular attention is given to characterization of the time-domain signal, which is essential in developing experimental trigger techniques. 
%The current availability of computational power sufficient for calculation of individual scattering amplitudes from $10^9$ shower particles on a personal computer makes such simulations feasible.
In what follows we will describe the particle-level model, and how it has been incorporated into the RadioScatter software package to simulate RF scattering from PSP.

\section{Particle-level PSP model}
\subsection{Derivation of the individual particle radiative contribution}

Our goal is to calculate the reflected radio-frequency signal due to the PSP, which requires, primarily, determining the individual particle equation of motion, and the properties of electic field wave propagation within the medium. 

 Our calculation starts from the classical equation of motion for an electron with label $A$, under the influence of an incident plane wave from a source at a distance $R_A$, and subject to collisions with frequency $\nu_c$, 

\begin{equation}\label{eom}
 m\left(\ddot{\mathbf{x}}_A+ \dot{\mathbf{x}}_A\nu_c\right) = -q\mathbf{E_0},%E_0e^{i(\mathbf{k} \cdot \mathbf{R_A} -\omega t)}\hat{\epsilon_A}
\end{equation}
with 

\begin{equation}
  \mathbf{E_0}=E_0e^{i(\mathbf{k} \cdot \mathbf{R_A} -\omega t)}\hat{\epsilon_A}.
\end{equation}

\begin{figure}[H]
\begin{centering}
\includegraphics[width=.8\textwidth]{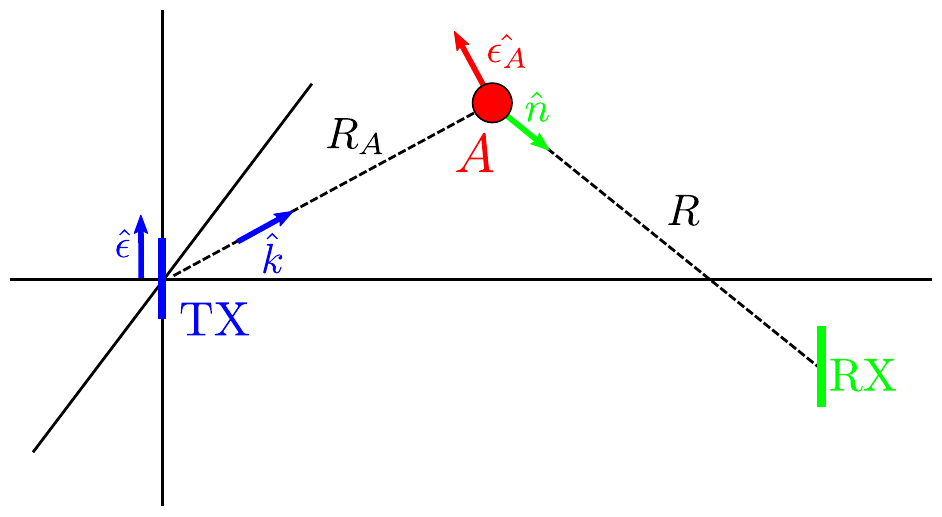}
\par\end{centering}
\caption{The angles used in the derivation of the invidual particle scattering contribution presented in the text. The direction of the wave vector $\hat{k}$ points from transmitter (TX) to the charge A. $\hat{n}$ points from the charge A to the receiver (RX). The polarization of the source is labeled $\hat{\epsilon}$, and the polarization of field at charge A is $\hat{\epsilon_A}$, which is perpindicular to $\hat{k}$ and lies in plane with $\hat{\epsilon}$.}
\label{scat_convent}
\end{figure}

The symbol $q$ is the electric charge. $\omega$ is the angular frequency of the source field. The unit vector $\hat{\epsilon_A}$ is the polarization vector of the source field as evaluated at the charge $A$, and $\mathbf{k}$ is the wave vector of the source electric field, and is complex. It will play an important role in calculations of the scattered field in what follows. A diagram of the angles is given in Figure~\ref{scat_convent}.

The collisional term $\nu_c$ is expressed as a sum over the species in the plasma, as in \cite{raizer}. 
\begin{equation}\label{collision}
 \nu_c = \sum_s n_s\bar{v_e}\sigma_s;
\end{equation}
here, $n_s$ and $\sigma_s$ are the number density and collisional cross-section, respectively, of species $s$, and $\overline{v_e}$ is the mean thermal velocity of the PSP electrons. (More discussion of collisions will follow in a later section.) Solving for the acceleration of the charge in Eq.~\ref{eom} gives
\begin{equation}
\ddot{\mathbf{x}}_A=-\frac{q\omega E_0e^{i(\mathbf{k} \cdot \mathbf{R_A} -\omega t)}}{m(\omega + i\nu_c)}\hat{\epsilon_A}.
\end{equation}
%where $\hat{\epsilon}$ is the polarization vector of the incident wave. %Inserting this into the Larmour equation for the electric field $\mathbf{E_a}$ of a single charge due to acceleration, and solving for the real part yields

For charges with negligible velocity, such as the ions and ionization electrons in the PSP cloud, the electric field from a charge A takes the form

\begin{equation}\label{efield}
\mathbf{E_A}=\frac{q}{R^2}\hat{n} + \frac{q}{c^2}\left[\frac{\hat{n}\times(\hat{n}\times\ddot{\mathbf{x}}_A)}{R}\right]_{ret},
\end{equation}
where the evaluation of the acceleration field takes place at the retarded time, $t^\prime=t-|\mathbf{R}|/c$ with $t$ the time at some distant receiver, and %In what follows, subscripts indicate the 4-space position at which a time is evaluated.
the unit vector $\hat{n}$ points from the charge to the receiving antenna. In the plasma approximation, the first term in Eq.~\ref{efield} cancels due to equal and opposite contributions from electrons and ions. However the second term is only nonzero for the free electrons, as the ions in a dense medium are fixed. So the problem reduces to calculating only the acceleration field of the free ionization electrons.

The far-field Larmor equation for the electric field of the charge A, under an incident field $\mathbf{E_0}$ and including the effects of collisions as above, is then
\begin{equation}\label{primary}
\mathbf{E_A}=-\frac{q^2\omega}{c^2m(\omega+i\nu_c)}\left[ \frac{\hat{n}\times(\hat{n}\times\mathbf{E_I})}{R}\right], 
\end{equation}
where 
% \begin{equation}
% \mathbf{E_I}=\frac{\mathbf{E_0}}{R_A}=\frac{E_0e^{i(kR_A-\omega t_A)}}{R_A}\mathbf{\hat{\epsilon_A}},
% \end{equation}
\begin{equation}
\mathbf{E_I}=\frac{V_0}{R_A}\mathbf{\hat{\epsilon_A}}=\frac{V_0e^{i(kR_A-\omega t_A)}}{R_A}\mathbf{\hat{\epsilon_A}},
\end{equation}
which is simply the incident field $\mathbf{E_0}$ at the point A. The quantity $V_0=E_0\times 1m$ is the source field evaluated 1 meter from the transmitting antenna, with units of Volts. %, where $R_A$ is the magnitude of the vector between the source and point A, and $t_A$ indicates that the phase is evaluated at the time the incident field reaches the charge A. 
For simplicity, it is assumed that the source $\mathbf{E_0}$ is plane polarized, %so the quantity $kR_A$ implies that
and the wave vector $\mathbf{k}$ lies along the vector $\mathbf{R_A}$. The charge acceleration vector $\mathbf{\hat{\epsilon_A}}$, which is the polarization vector of the source at A, forms a plane with the polarization vector of the souce, perpendicular to $\mathbf{R_A}$. The quantity R (without subscript) is the magnitude of the vector between the charge A and the receiver (Figure~\ref{scat_convent}).

%new wave vector crap
When dealing with the propagation of waves in a dense medium, the properties of the medium itself must be considered. For a general treatment of the radar problem, there are three propagation regions for RF wave numbers $k$: free-space, in-medium, and in-plasma, which we denote as 
\begin{alignat}{2}
  k_0&=\text{k}_0,\\
k_m&=\text{k}_m&&-i\xi(\omega),\quad\text{and}\\
k_p&=\text{k}_p&&-i\beta,
\end{alignat}
% \begin{alignat}{3}
%   k_0&=\frac{\omega}{c}&&=\text{k}_0,\\
% k_m&=\frac{\omega}{c}\text{n}_m&&=\text{k}_m&&-i\xi(\omega),\quad\text{and}\\
% k_p&=\frac{\omega}{c}\text{n}_p&&=\text{k}_p&&-i\beta,
% \end{alignat}
respectively. The non-italicized k represents the real part of $k$, $\xi(\omega)$ is the frequency-dependent attenuation coefficient of the medium, and $\beta$ is the attenuation coefficient of the plasma due to collisions (discussed below). For the medium, $\xi(\omega)$ is the inverse of the attenuation length, a quantity representing the length over which a field amplitude is reduced by a factor of $e$ \cite{mark_attn}. 
For the ionization electrons in a plasma subject to the equation of motion of Eq.~\ref{eom}, the complex wave number $k_p$ is \cite{jackson},

%When dealing with the propagation of waves in a dense medium, the properties of the medium itself must be considered. For the ionization electrons in a material subject to the equation of motion of Eq.~\ref{eom}, the complex wave number $k$ is \cite{jackson},

\begin{align}
k_p=\frac{\omega}{c}\text{n}_p&= \frac{\omega}{c}\left[1-\frac{4\pi n_eq^2}{m}\left(\frac{1}{\omega^2+i\omega\nu_c}\right)\right]^{\frac{1}{2}}\\
&\approx \frac{\omega}{c}\left[1-\frac{4\pi n_eq^2}{2m}\left(\frac{1}{\omega^2+i\omega\nu_c}\right)\right]\\
&\approx \frac{\omega}{c}\left[1-\frac{\omega_p^2}{2}\left(\frac{1}{\omega^2+i\omega\nu_c}\right)\right],
\end{align}
where we have used the binomial approximation for the index of refraction $\text{n}_p$, and have introduced the plasma frequency, $\omega_p=\sqrt{4\pi n_eq^2/m}$, where $n_e$ is the electron number density in units of $cm^{-3}$. This number density is the local number density at the charge A. The imaginary part of this expression represents a damping of wave propagation in the plasma due to collisions,

\begin{equation}
  \label{eq:impart}
\beta=  \text{Im}[k_p]\approx\frac{\omega_p^2}{2c}\left(\frac{\nu_c}{\omega^2+\nu_c^2}\right).
\end{equation}

\begin{figure}[H]
\begin{centering}
\includegraphics[angle=-90,width=.8\textwidth]{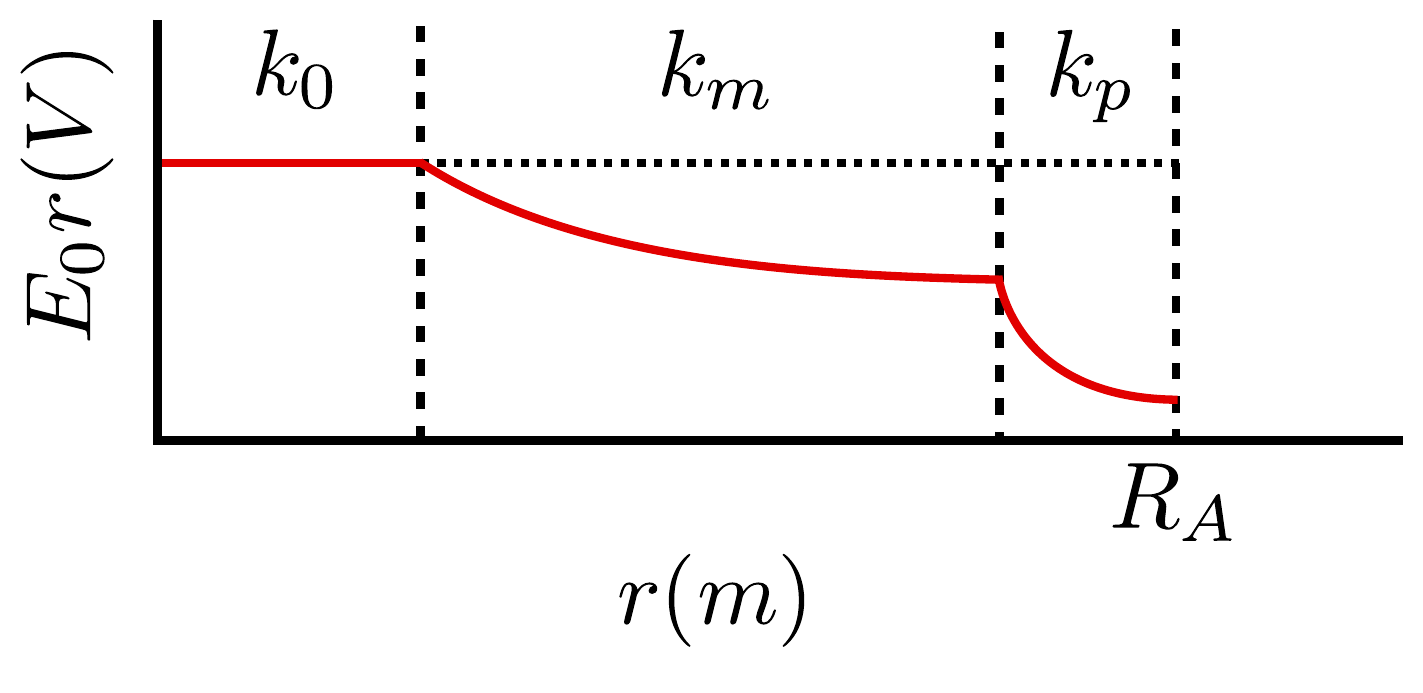}
\par\end{centering}
\caption{Cartoon representing the amplitude of a wave (solid line) as it propagates through mediums with different n and $k$.}
\label{k_vec}
\end{figure}

Continuing with our general assumption that a transmitter may be in free space interrogating a plasma within a dense medium, we must expand the quantity $kR_A$ as a sum over the different regions and their associated wave vectors, shown visually in Figure~\ref{k_vec}.

The general distance between TX and the charge A, $R_A$, is broken down into the same three regions as for $k$ above. %The boundary between free space and the medium is $R_m$, and $R_p$ is the boundary between the medium and the plasma. 
The distances travelled in free space, the medium, and the plasma, are $R_0$, $R_m$, and $R_p$ respectively, such that $R_A=R_0+R_m+R_p$.
We then expand the incident field $\mathbf{E_I}$ using these regions.

\begin{align}
\mathbf{E_I}&=\frac{V_0}{R_A} \text{exp}\left[i(kR_A - \omega t)\right]\mathbf{\hat{\epsilon_A}}\\
%&=\frac{V_0}{R_A} \text{exp}\left[i(k_0R_m + k_m(R_p-R_m) + k_p(R_A-R_p) - \omega t)\right]\\
&=\frac{V_0}{R_A} \text{exp}\left[i(k_0R_0+k_mR_{m}+k_pR_{p} - \omega t)\right]\mathbf{\hat{\epsilon_A}}\\
&=\frac{V_0}{R_A} \text{exp}\left[i(\text{k}_0R_0+\text{k}_mR_{m}+\text{k}_pR_{p} - \omega t)\right] e^{-\xi R_{m}}e^{-\beta R_{p}}\mathbf{\hat{\epsilon_A}}
\end{align}

Because $k_p=k_p(\omega_p)$, $\omega_p=\omega_p(n_e)$, and $n_e=n_e(R_p)$, the product of the wave vector (real and imaginary) with the plasma path length $R_p$ is in fact an integral over the distance $R_p$. For example, the damping term from the imaginary part of the wave vector, $\beta$, becomes

\begin{equation}
  \exp\left[-\beta R_p\right] \rightarrow \exp\left[-\int_0^{R_p} \beta(r) dr\right],
\end{equation}
which accounts for the variation in $k_p$ as it traverses the plasma.\footnote{This integration is currently neglected in the RadioScatter module, due to the computational expense involved. However, at the energies/densities/frequencies of interest to this problem ($10^8$-$10^9$Hz, $10^{15}$-$10^{21}$eV) the single-value approximation is acceptable. For example, at 500~MHz and 100PeV, the difference in attenuation of an incident wave between the single-value and integral method is $<$10\%, with the single-value method being the more conservative, in terms of the strength of the signal returned. A future release will include the effects of this integration.}

Altogether, we now have an expression for the scattered field from an individual electron in the PSP cloud, 

\begin{equation}
  \label{eq:pretty}
  \mathbf{E_A}=\frac{\alpha E_I}{R}\hat{n}\times\hat{n}\times \hat{\mathbf{\epsilon_A}}
\end{equation}
where
\begin{equation}
  \alpha=-\frac{q^2\omega}{c^2m(\omega+i\nu_c)}
\end{equation}
is complex with units of length. When the imaginary part of $\alpha$ goes to zero, that is, when $\nu_c=0$ and there are no collisions, $\alpha$ is the familar `classical electron radius', and the collisional damping term $\beta$ goes to zero. In this case, %$k_p=c^{-1}(\omega^2-\omega_p^2)^{1/2}$, %the plasma wave number $k_p$, 
\begin{align}
\label{real_wavenumber}
% %\text{Re}[k]&=\text{Re}\left[\frac{\omega}{c}\text{n}\right]\\
% k_p&=\frac{\omega}{c}\text{n}_p\\
% &=\frac{\omega}{c}\left(1-\frac{\omega_p^2}{\omega^2}\right)^{\frac{1}{2}}\\
k_p &=\frac{1}{c}\sqrt{\omega^2-\omega_p^2},
\end{align}
which is the standard dispersion relation for electric fields in a collisionless plasma.

 \section{Applicability}
 The parameters $\alpha$ and $\beta$ must be experimentally verified, as $\nu_c$ is not known for the case of a particle shower in ice. We can however use standard plasma theory\cite{nicholson} and also experimental data\cite{meteor1} to assess the validity of the above model. This single-particle expression is applied to all particles within a shower to attain the full scattered signal. Details about the sum, which incorporates the other main unknown of the model, the plasma lifetime $\tau$, are given in a later section. What follows in this section pertains to the sum total scattered signal from a shower.

The characteristics of scattered RF, with angular frequency $\omega$, reflected from a plasma are determined by the magnitude of $\omega$ relative to $\omega_p=\omega_p(n_e)$.
For regions of high electron number density where $\omega<\omega_p$ (``overdense" regime), the wavenumber of Eq.~\ref{real_wavenumber} (collisionless regime) is fully imaginary, and therefore that region is opaque to incident RF, e.g. these fields are fully reflected. For $\omega>\omega_p$ (``underdense"), reflection is primarily dur to Thomson scattering and the plasma is increasingly transparent. For example, the very diffuse plasma in the Earth's ionosphere is traversed with minimal scattering loss by ultra-high-frequency (UHF) RF transmissions from satellites (underdense regime), whereas low-frequency waves broadcast from Earth may be totally reflected (overdense regime).

\begin{figure}[H]
\begin{centering}
\includegraphics[width=.8\textwidth]{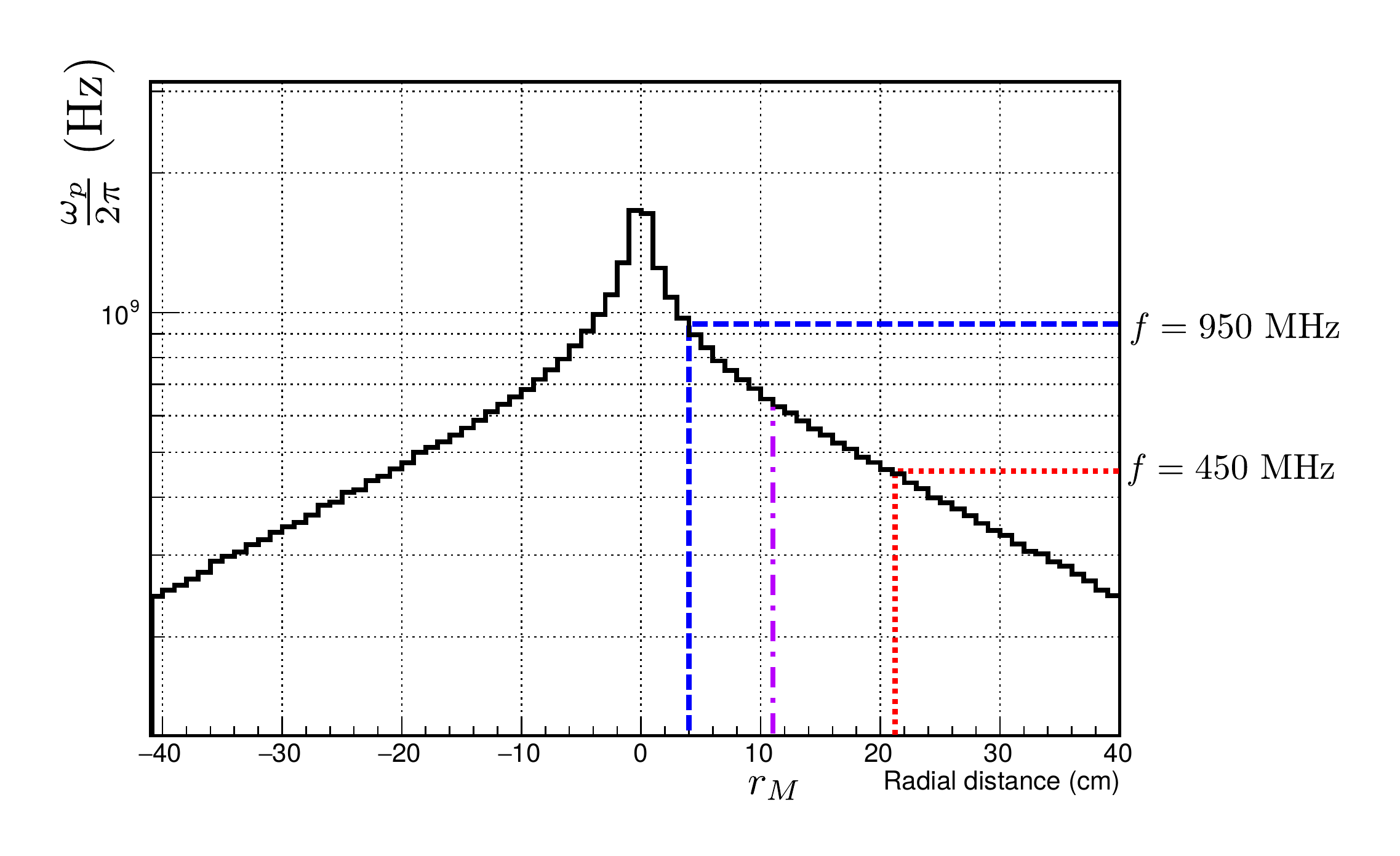}
\par\end{centering}
\caption{Shower plasma frequency $\omega_p/2\pi$ versus shower lateral profile for a 10~PeV shower. The effective corresponding radial extent of the overdense region is also shown for two interrogating frequencies. ``$r_M$" denotes the Moliere radius in ice.}
\label{plasma_freq}
\end{figure}

Therefore, in general, overdense scattering is coherent and underdense scattering is incoherent, and so radar sounding is descibed in terms of overdense scattering. The effective cross-section of the overdense region of a generic radar target can be calculated from the standard bi-static radar equation, as follows:
% %It is a further fact that the received power in the overdense case can be used to calculate an effective cross-sectional surface from which the transmitted RF has scattered. This is of fundamental importance in bi-static RADAR sounding. We can calculate the effective scattering cross-section $\sigma_{eff}$ by solving for it using the standard bi-static radar equation:
 \begin{equation}\label{rcs}
  \sigma_{eff}=\frac{(4\pi)^3 R_t^2 R_r^2 P_r}{P_t G_t G_r \lambda^2},
 \end{equation}
 where $R_t$ and $R_r$ are the distances from the shower to the transmitter and receiver, respectively, $P_t$ and $G_t$ are the transmitted power and transmitter antenna gain, and $P_r$ and $G_r$ are the received power and receiver antenna gain. %$\sigma_{eff}$ provides an invariant measure for scattering efficiency with respect to $\omega$, $R_r$, and $P_t$, because the size of the overdense region goes as $\lambda^2$, and for a fixed $R_t$, $P_r$ goes as $R_r^{-2}$. 
 In the case of a particle shower, $\sigma_{eff}$ is bounded by the product of the transverse scale (of order the Moliere radius, or $\cal{O}$(10 cm) for ice) and the longitudinal scale (set by the radiation length, or $\cal{O}$(10 m) for ice) of the reflecting shower. In our case, the overdense/underdense boundary, in addition to being frequency-dependent, is evolving both spatially, over distances of cm, and temporally, over times of order ns. The spatial dependence on interrogating frequency $f=\omega/2\pi$ is shown diagrammatically in Figure~\ref{plasma_freq}, where $\omega_p/2\pi$ is plotted versus lateral profile for a 10~PeV shower. The x-intercepts indicate the lateral extent of the overdense region for different sounding frequencies, and the Moliere radius $r_M$ is also indicated; the greater penetration of the higher-frequency signal is evident from the Figure. %Note that, for primary particle energies below this, the scattering is purely undersense for these frequencies, as they will be above the plasma frequen%For any frequency, $\sigma_{eff}$ cannot exceed the physical extent of the shower, giving us a constraint.

% \begin{figure}[H]
% \begin{centering}
% \includegraphics{k_vec}
% \par\end{centering}
% \caption{Diagram of the amplitude of a wave as it propagates through mediums with different $k$.}
% \label{plasma_freq}
% \end{figure}

\begin{figure}[H]
\begin{centering}
\includegraphics[width=.8\textwidth]{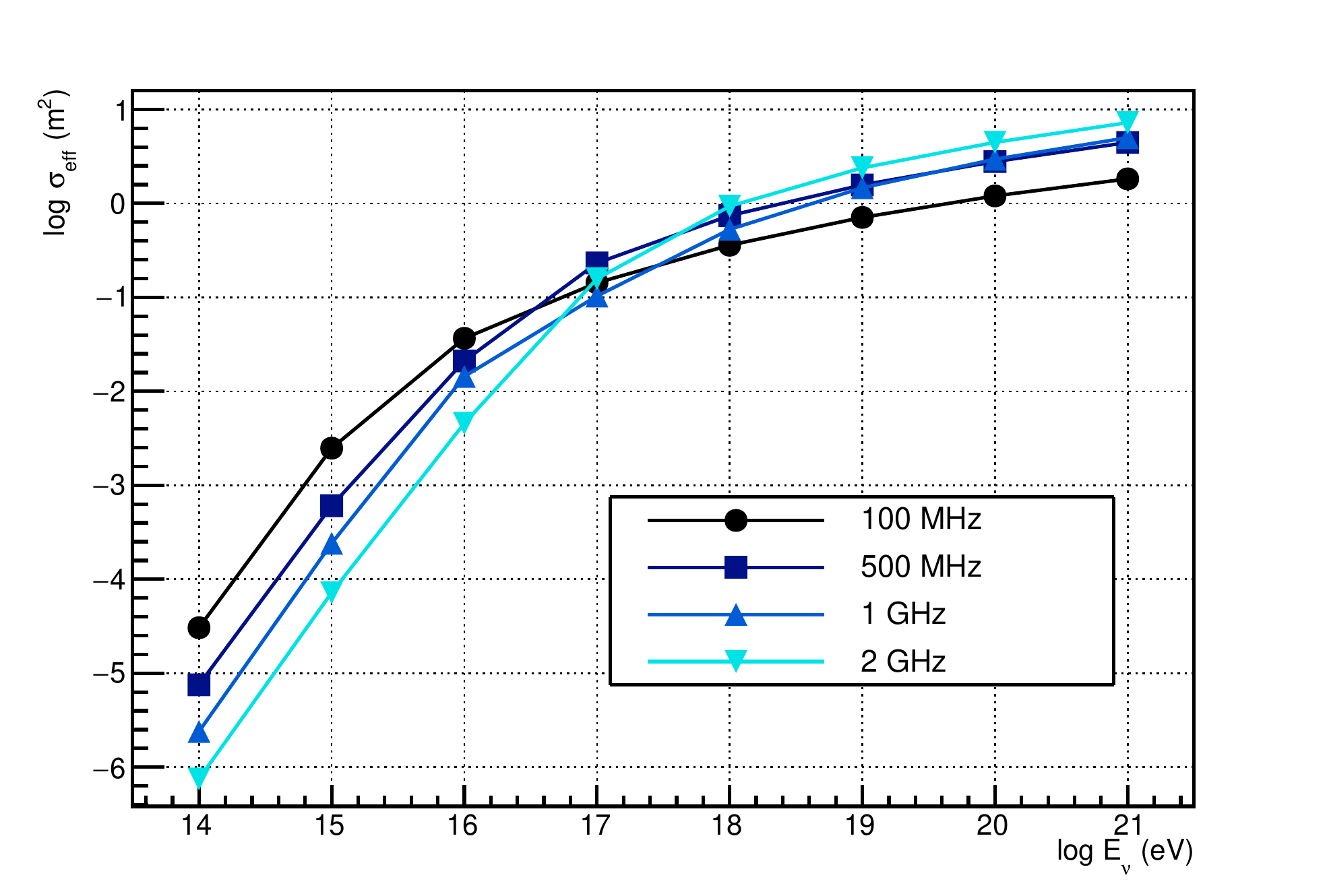}
\par\end{centering}
\caption{Effective scattering cross-section $\sigma_{eff}$ as a function of primary energy, for a range of transmitted frequencies. The transmitter output is 1kW ($\sim$223~V) and the TX--RX baseline is $\sim$1~km, with showers thrown at random positions within the intervening volume. The plasma lifetime is 1~ns.}
\label{sigma_eff_v_freq}
\end{figure}

We can use this reasonable upper bound on $\sigma_{eff}$ to assess the validity of Eq.~\ref{eq:pretty}, by plotting $\sigma_{eff}$ versus primary particle energy for various interrogating frequencies. Figure~\ref{sigma_eff_v_freq} shows that $\sigma_{eff}$ remains reasonable (e.g. on the order of the dimensions described above) up to high energies, accross a wide range of frequencies. %Evident from the figure is the underdense/overdense transition beginning at $\sim$10~PeV energies, where the cross section begins to level off.  
From a macroscopic standpoint, lower interrogating frequencies see a larger physical cross section of the shower (due to the plasma frequency) than high frequencies, but ultimately smaller $\sigma_{eff}$ than high frequencies at high energies due to the $\lambda^{-2}$ term in Eq.~\ref{rcs}. We see this same behavior in the particle level treatment (Figure~\ref{sigma_eff_v_freq}), where $\sigma_{eff}$ scales with frequency once the overdense/underdense boundary is crossed ($\sim$1-10~PeV).

% We hope to extract the parameter $\beta$ from data taken in the upcoming test beam experiment and thereby validate the mathematical approach presented herein.%For the spatial and temp

\section{RadioScatter}
The above model is incorporated into a software package called RadioScatter~\cite{radioscatter}, which is open source and has been successfully run on several different flavors of linux. The module is written in C++ and can be incorporated into user scripts or large Monte-Carlo packages such as GEANT4. The code, documentation, and example GEANT4 programs using RadioScatter are available at the referenced GitHub repository.

\begin{figure}[H]
\begin{centering}
\includegraphics[width=.7\textwidth]{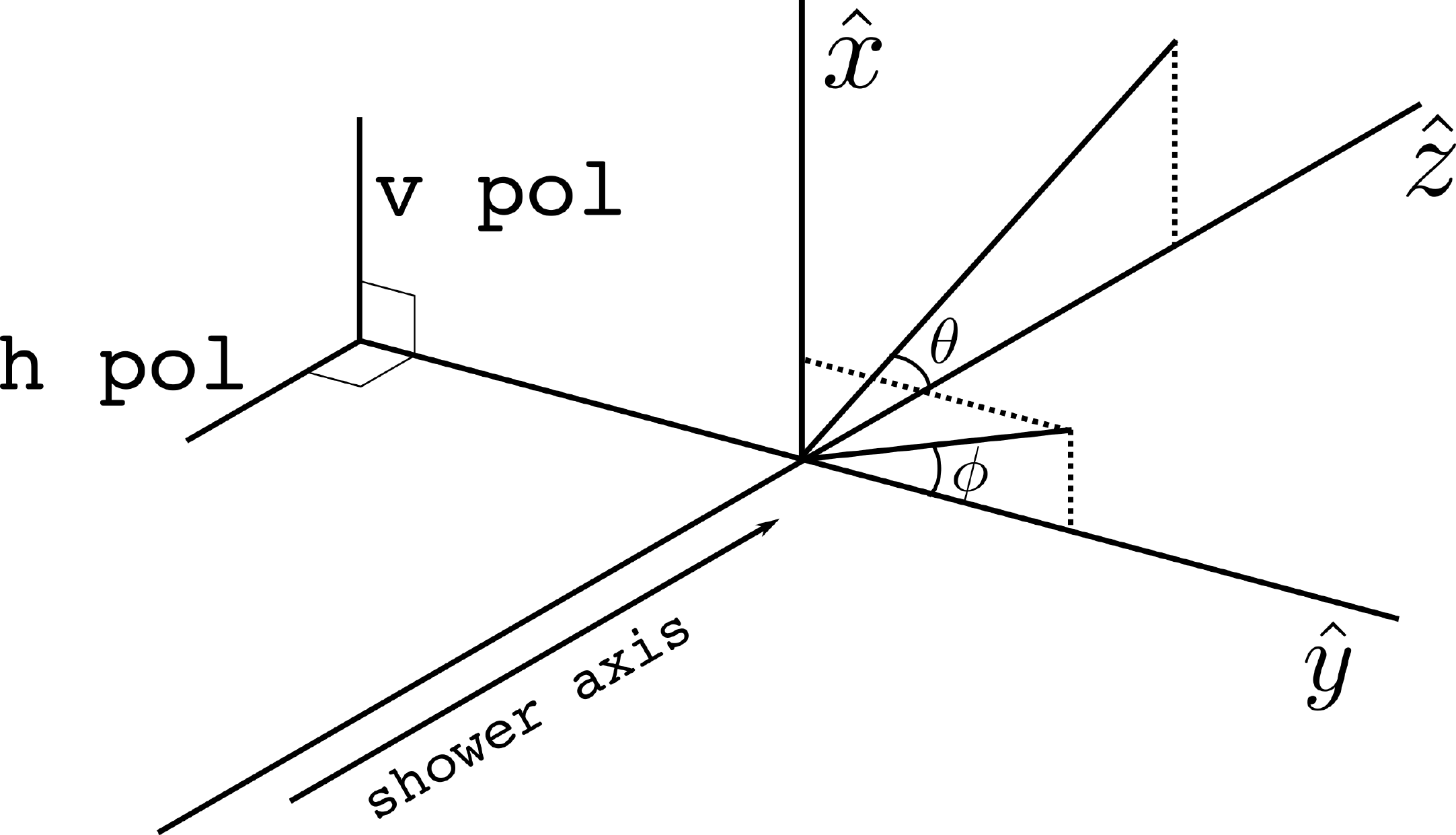}
\par\end{centering}
\caption{Geometry conventions used in RadioScatter, indicating what is meant by vertically (v pol) and horizontally (h pol) polarized antenna configurations used in the text and the module.}
\label{conventions}
\end{figure}

The polarization and angle conventions used in RadioScatter are presented graphically in Figure~\ref{conventions}.

\section{GEANT4 implementation}
We now describe the actual implementation of RadioScatter within the GEANT4 simulation package. We describe how the PSP is generated, how number densities and collision frequencies are calculated, and the technique for calculating the scattered signal from the PSP.

\subsection{Generation of the PSP}
GEANT4 is the premier suite of simulation tools for particle interactions with matter. Users can specify nearly any projectile incident on nearly any target material and geometry, with access to individual four-momenta at run-time.
GEANT4 provides this particle-level information to the user at each step of a shower's evolution, including the length of each step in mm (medium-density specific, and internally-defined in GEANT4) and the energy deposited in the medium over that step. GEANT4 utilizes an extensive library of materials and their properties, including radiation lengths and ionization energies. To find the number $N$ of ionization electrons produced in each step of each shower particle, we therefore divide the amount of energy deposited in the step by the ionization energy of the medium. In ice, for example, GEANT4 calculates an ionization energy of 69~eV. %, and round down to the nearest integer number.
It is these ionization electrons which comprise the PSP cloud and from which we calculate the scattered signal.

\begin{figure}[H]
\begin{centering}
\includegraphics[width=\textwidth]{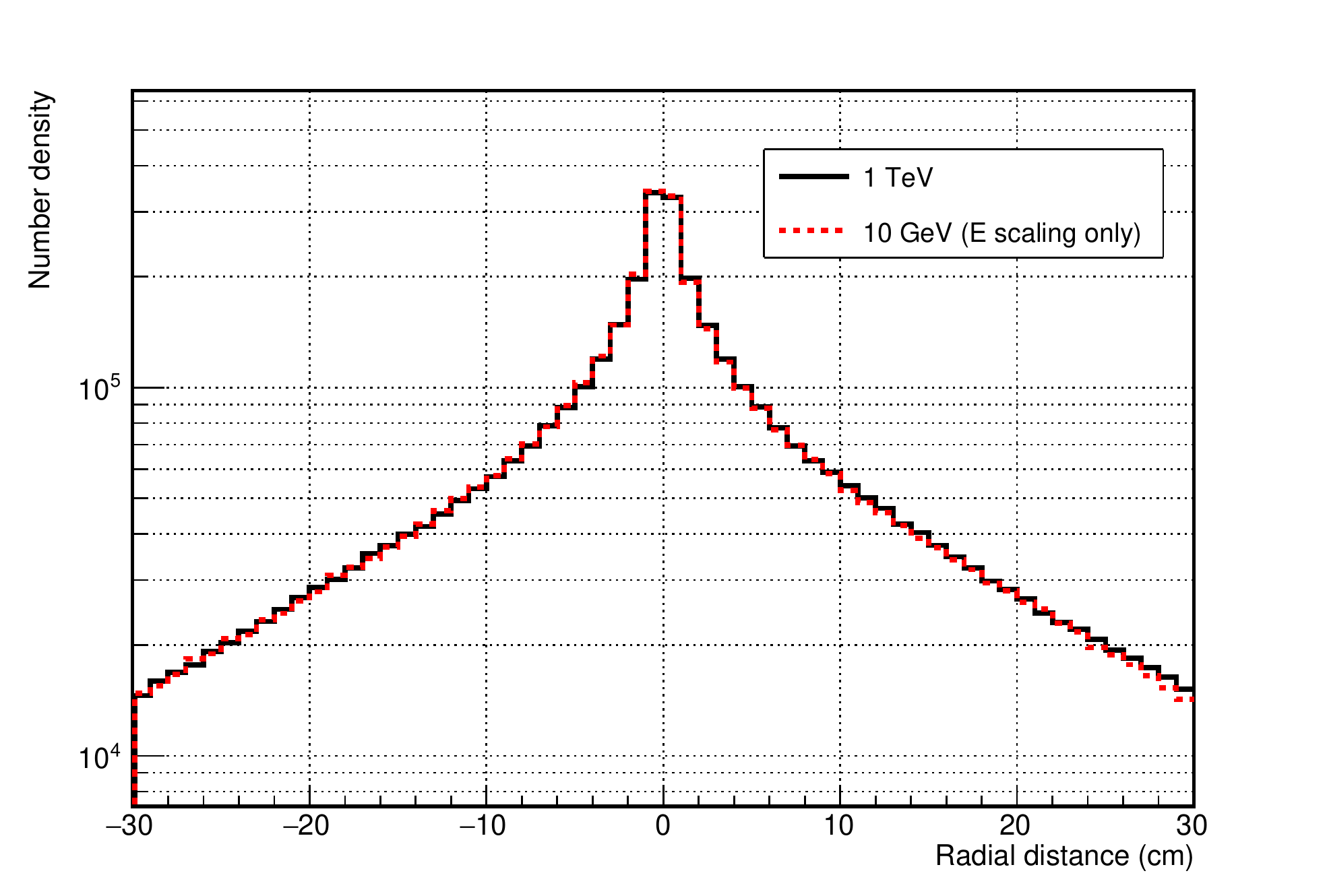}
\par\end{centering}
\caption{Example radial distribution of shower particles in GEANT4, showing the invariance in lateral distribution as a function of primary particle energy. The 10~GeV shower has been scaled by a factor of 100 and overlaid with an unscaled 1~TeV shower profile.}
\label{radial}
\end{figure} 

GEANT4 can produce showers on a personal computer at energies up to roughly 10~TeV, but beyond that, it becomes computationally inefficient to produce a large sample. Therefore, in order to efficiently produce showers at higher energies in large numbers, simple linear scaling is applied, both in the longitudinal direction and in time, to showers of lower energies. To calculate the correct scaling factors, numerous GEANT4 showers were produced at decades of primary particle energy from 100~MeV up to 10~TeV and analyzed. 

\begin{figure}[H]
\begin{centering}
\includegraphics[width=\textwidth]{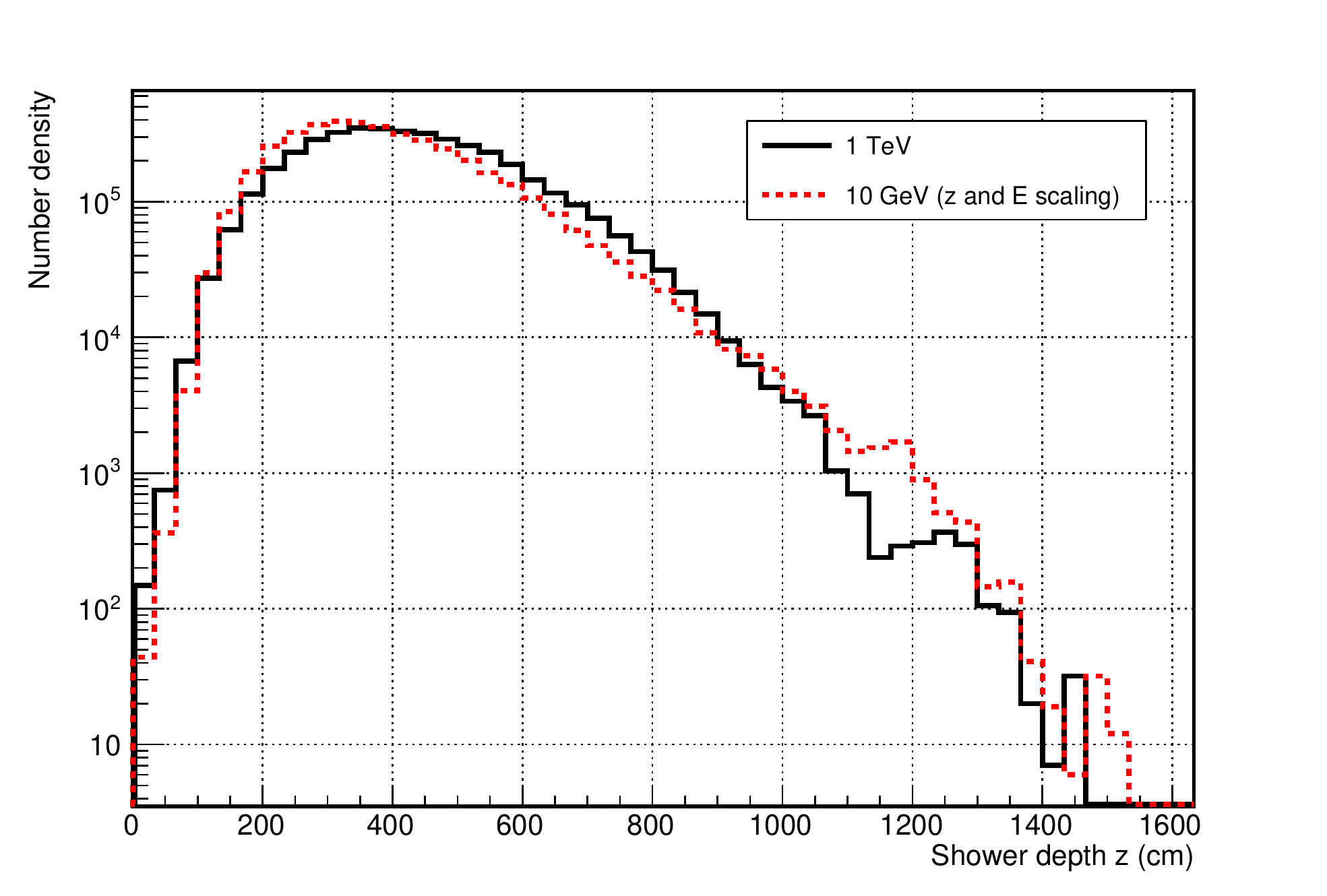}
\par\end{centering}
\caption{Longitudinal distribution of showers in GEANT4. A 10~GeV shower has been scaled in number density and the longitudinal dimension to match a higher energy, 1~TeV shower.}
\label{scaling}
\end{figure}

The radial distribution for a shower in a medium is largely independent of primary energy, with 90\% of the particles contained within 1 Moliere radius, which for ice is $\sim$10~cm. This is shown in Figure~\ref{radial}, where a 10~GeV shower has been scaled by number density only, which makes the lateral shower profile match a 1~TeV shower profile. The longitudinal length of the shower scales with the log of the primary energy, as does the shower duration. Therefore we apply the proper scalings to the longitudinal and time components for each ionization 4-vector in the shower for a target primary energy, and scale the number density accordingly.  This results in shower profiles which mimic those at energies beyond what is accessible in GEANT4. A comparison of a 1~TeV shower with a 10~GeV shower that has been scaled up is shown in Figure~\ref{scaling}, showing good agreement in longitudinal profile. For computational efficiency, a scaled 10~GeV shower is used in the RadioScatter module for all higher energies. While this is clearly not an ideal description of shower shape at very high energies, this technique allows for a reasonable approximation for the purposes of this simulation.  We note that the longitudinal scaling factor required to scale a 10~GeV shower length up to that of a 1~EeV shower is $\sim$4, so the maximum scaling is overall less than an order of magnitude. The length and time scaling can be turned on and off by the user in RadioScatter.

We note that RadioScatter neglects the Landau-Pomeranchuk-Migdal (LPM) effect\cite{lpm_1}\cite{lpm_2} in the longitudinal shower profile. This effect, detailed for the radio problem in \cite{lpm_3} and \cite{lpm_4}, is a suppression of low-energy bremsstrahlung and pair production in showers at very high energies, resulting in an effective lengthening of showers in the longitudinal dimension. This effect would have minimal impact on the radar problem as the extended tail of the distribution at thigh energies has a low number density relative to shower maximum, and therefore will not be part of the overdense scattering discussed above. 

\subsection{Calculation of the signal}

From the 4-vectors of ionization electrons provided by the GEANT4 simulation, we calculate the scattered fields for a specified interrogation frequency using the real part of Eq.~\ref{eq:pretty}. %From the ionization 4-vector, we determine the retarded time at the transmitter, and, from that, the phase of the incident wave at the point of interaction.   
The resultant fields for all PSP particles are propagated back to the receiver and summed in time bins corresponding to the user-defined sampling period. For example, the resultant real part of the total electric field at the receiver for a single sampling period $T$ is given by

\begin{equation}
 Re\left[\mathbf{E}_{tot}\right]=\frac{1}{T}\sum_{n=1}^{N}\int^{t+T}_{t}\Theta(t^\prime-t^i_n)\Theta(t^f_n-t^\prime)Re\left[\mathbf{E}_n(t)\right]dt,
 \end{equation} \label{calc}
where $\mathbf{E}_n$ is given in Eq.~\ref{eq:pretty}. $t^\prime=(t-|\mathbf{R}|/c)$ is the retarded time at the position of charge $n$, and the step functions ensure that the charge $n$ exists at the retarded time, with $t^i_n$ and $t^f_n$ being the production and recombination/attachment (initial/final) times, respectively, for charge $n$. These are a function of the plasma lifetime $\tau$. The factor $1/T$ is present because, in practice, a standard digitizer effectively averages the measured voltage over the sampling period, so we similarly calculate the average value of each $\mathbf{E}_n$ over a single sampling period, in order that the displayed voltage will be independent of the time base, and sum these average values. We then take this electric field $\mathbf{E}_{tot}$ and multiply by an antenna effective length to obtain, e.g. the voltage read on an oscilloscope. %The imaginary portion, which defines attenuation, is calculated, as well.

 \subsection{Example signal}

\begin{figure}[H]
\begin{centering}
\includegraphics[width=\textwidth]{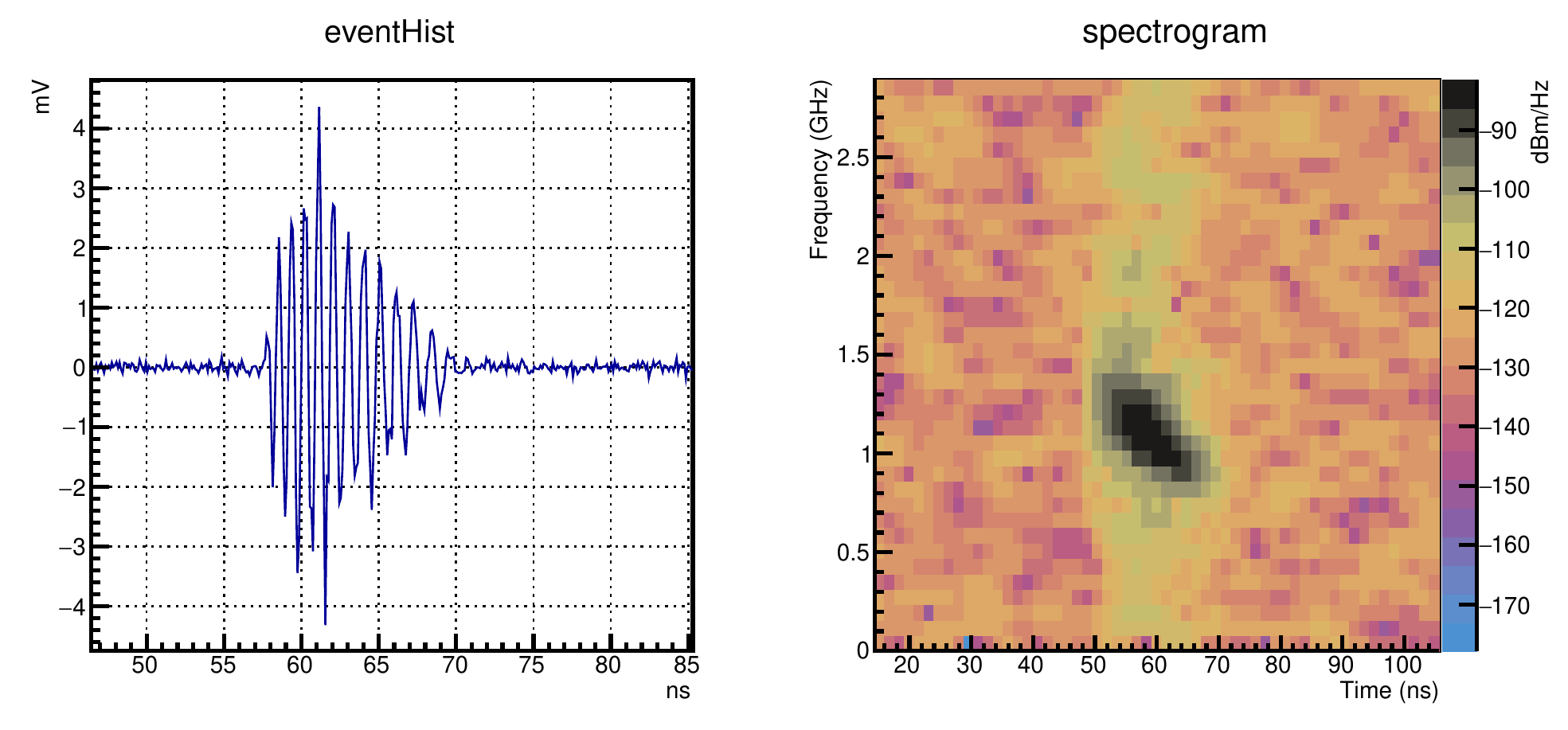}
\par\end{centering}
\caption{Simulated radio reflection for a 5~GHz bandwidth receiver, from an electron-initiated plasma consisting of $10^9$ 13.6~GeV primaries, superimposed upon thermal noise, with a sounding frequency of 1.15~GHz CW. The transmitter output power is 10~W and the plasma lifetime is 0.1~ns. The observed chirp-like signal is a function of the TX-PSP-RX geometry.}
\label{sig}
\end{figure}

Figure~\ref{sig} is an example of a simulated reflection from a GEANT4 shower using RadioScatter, where we have used Eq.~\ref{calc} to build up a time-domain signal. In this simulation, a 13.6~GeV electron beam with a bunch count of $10^9$ electrons (scaled per the above discussion, to the parameters of our upcoming SLAC testbeam, discussed below) is incident on high-density polyethylene (HDPE). The target is interrogated with 1.15~GHz continuous-wave (CW) radio signal at 100~mW output power, with horizontally polarized (i.e., antennas in the same plane as the shower axis) TX and RX. The plasma lifetime is set at 0.1~ns, and will be discussed further below. 

The `chirp' signal  of Figure~\ref{sig} is a function of the TX/RX proximity to the shower in this test-beam setup. For geometries where the TX-PSP baseline is much greater than the length of the shower itself, the `chirp' is replaced by a CW return at a shifted frequency away from the carrier. Experimentally, such a unique signal can be used to advantage in a low signal-to-noise trigger, as in \cite{sammy}\cite{firmware_trig}. The phase relationships between reflections from different parts of the plasma as it progresses through 4-space result in a coherent frequency shift of the received signal, even though none of the scatterers themselves have any appreciable 3-velocity, and the interrogating radio is monochromatic. This shift, observed in both the horizontal and vertical polarizations, is a function of the TX-PSP-RX geometry, and can be used to deduce position and direction information of the primary particle. Detailed analysis of the frequency shift/geometry relationship will be elucidated in a forthcoming article. 

\begin{figure}[H]
\begin{centering}
\includegraphics[width=\textwidth]{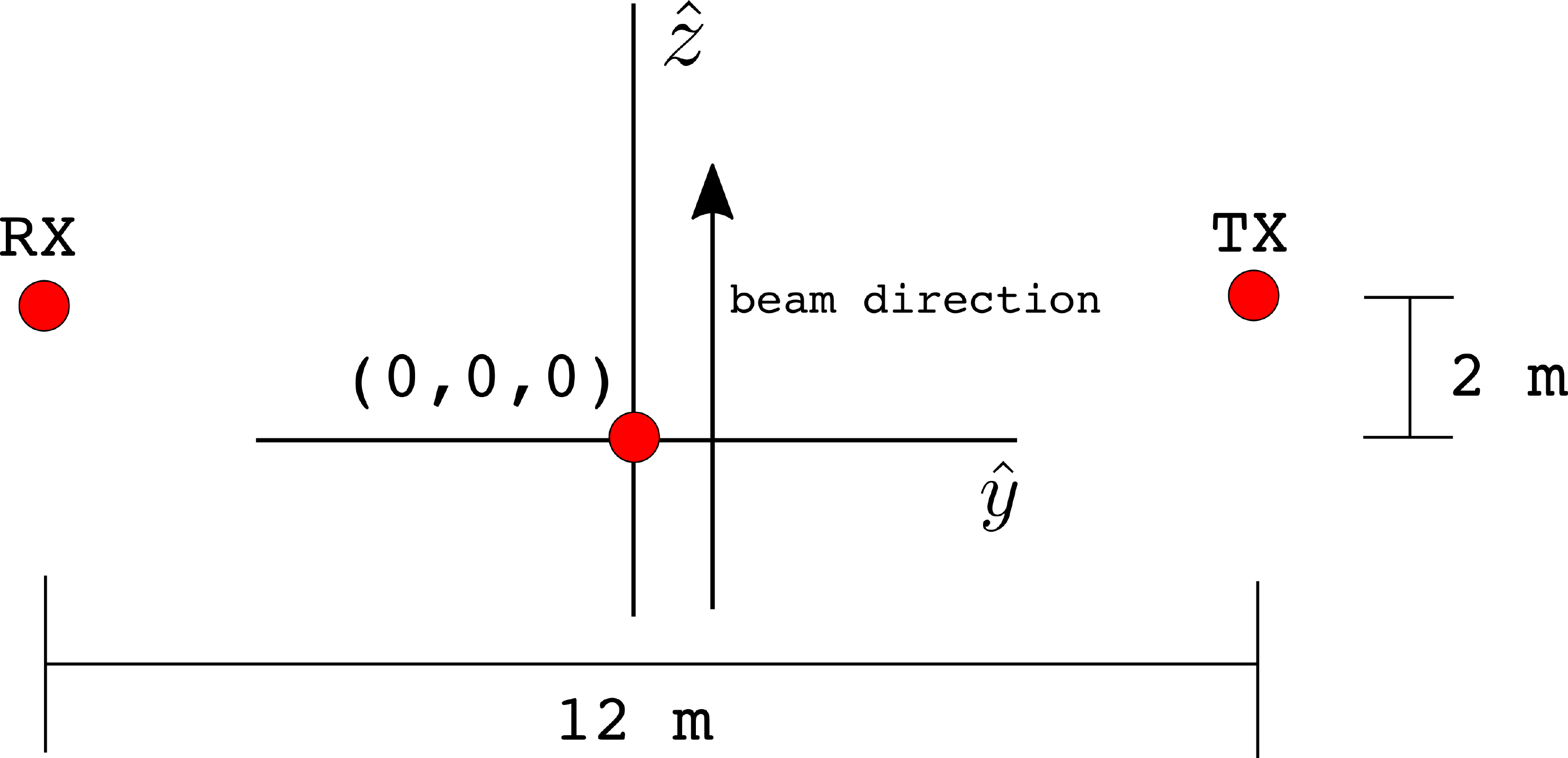}
\par\end{centering}
\caption{The geometry of the radar set-up for Figure~\ref{sig}. The shower vertex is at (0,0,0) with the shower progressing in the $+z$ direction. }
\label{example_geometry}
\end{figure}

The TX--RX--PSP geometry for this event is shown in Figure~\ref{example_geometry}. In this example, the coordinate system is set so that the shower vertex occurs at (0,0,0) and the shower evolves in the $+\hat{z}$ direction.  

\subsection{Plasma lifetime}
\begin{figure}[H]
\begin{centering}
\includegraphics[width=\textwidth/2]{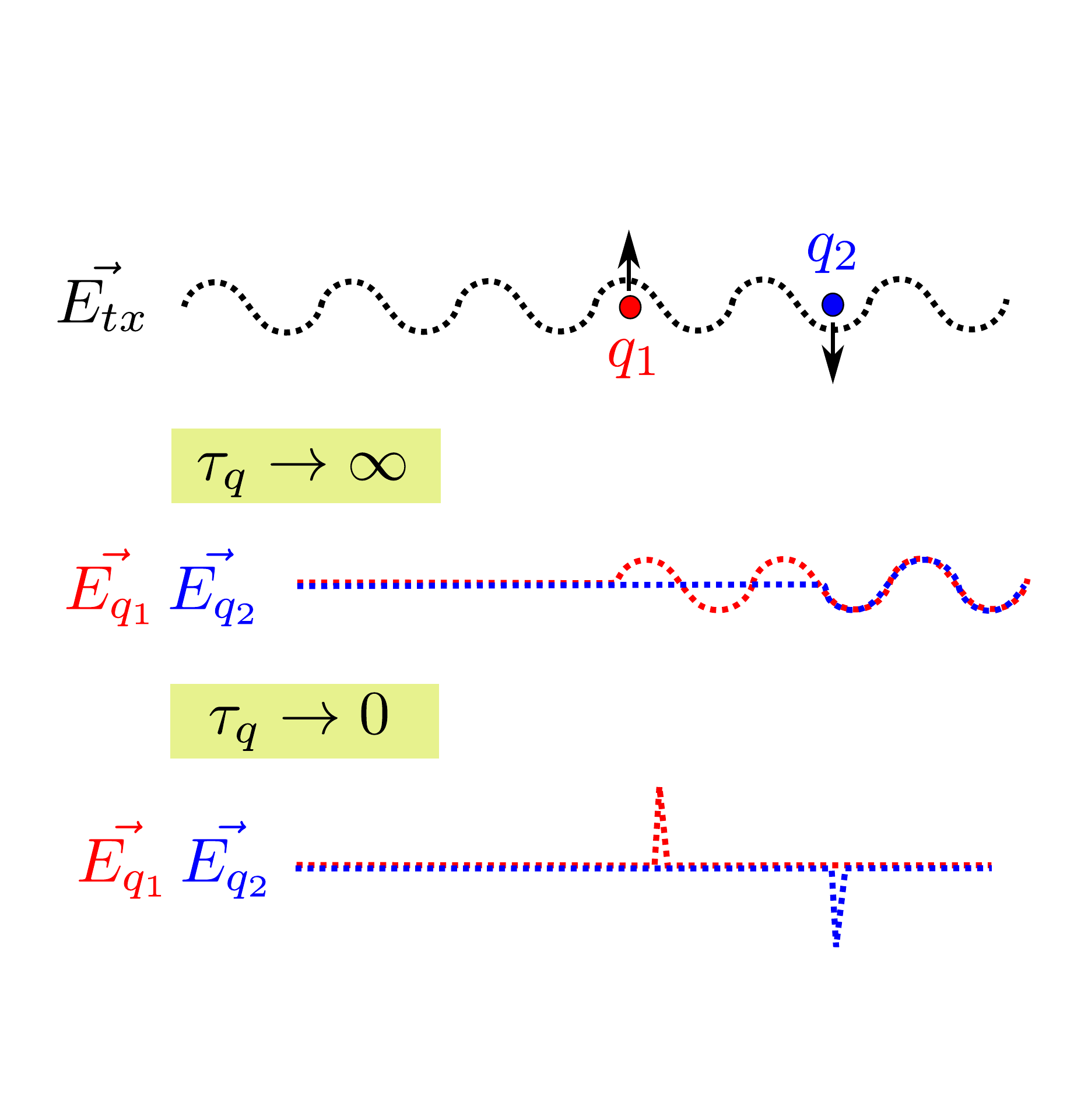}
\par\end{centering}
\caption{Graphical representation of the limiting cases for the free electron lifetime $\tau$. As $\tau\rightarrow\infty$, once the charges $q_1$ and $q_2$ are freed from the medium, they begin to radiate in phase with the incident field. As $\tau\rightarrow 0$, the charges move only briefly (less than a single oscillation period), and their polarity is given by the instantaneous phase of the incident RF. }
\label{scattering_concept}
\end{figure}

A primary unknown in the PSP problem is the true plasma lifetime $\tau$ for a given material, presumably dominated by ionic recombination or attachment to neutrals. %In RadioScatter, we have conservatively assumed zero lifetime. That is, the incoming fields are incident upon a single point in 4-space, which then instantaneously re-radiates to the receiver. 
In the classical picture, a free charge will oscillate in phase (or directly out-of-phase, if the charge is negative) with an incident field. In the limit that $\tau$ for this charge approaches zero, that is, $\tau << 1/f$, where $f$ is the interrogation frequency, the charge does not ``live'' long enough to make a full oscillation. Instead, the charge gets a `kick' from the field, with a direction dictated by the polarization and phase of the incident RF at that point in 4-space\cite{krijn_personal}.\footnote{This `kick' is due to the interrogating field. We assume the electron pops into stationary existence upon ionization, gets a kick from the field, and pops out of existence upon attachment or recombination. It is assumed that the start and end points of this process result in negligible RF emission, due to the non-relativistic velocities involved.}  %This `kick' is akin to a delta function in time space, reminiscent of the time-domain treatment of Cherenkov radiation in \cite{zhs}. 
The different time-scales are compared graphically in Figure \ref{scattering_concept}. 
For this reason, we expect to see coherent scattering even for lifetimes well below the period of an interrogating wave, since the individual, short-time kicks are correlated, being functions of the incident wave. And indeed, though the amplitudes are diminished, coherent scattered signals are seen in the simulation at lifetimes as short as 100~ps. 
%Note that, experimentally, the lifetime may be (unmeasurably) shorter than typical laboratory measurement signal sampling timescales.

\begin{figure}[H]
\begin{centering}
\includegraphics[width=\textwidth]{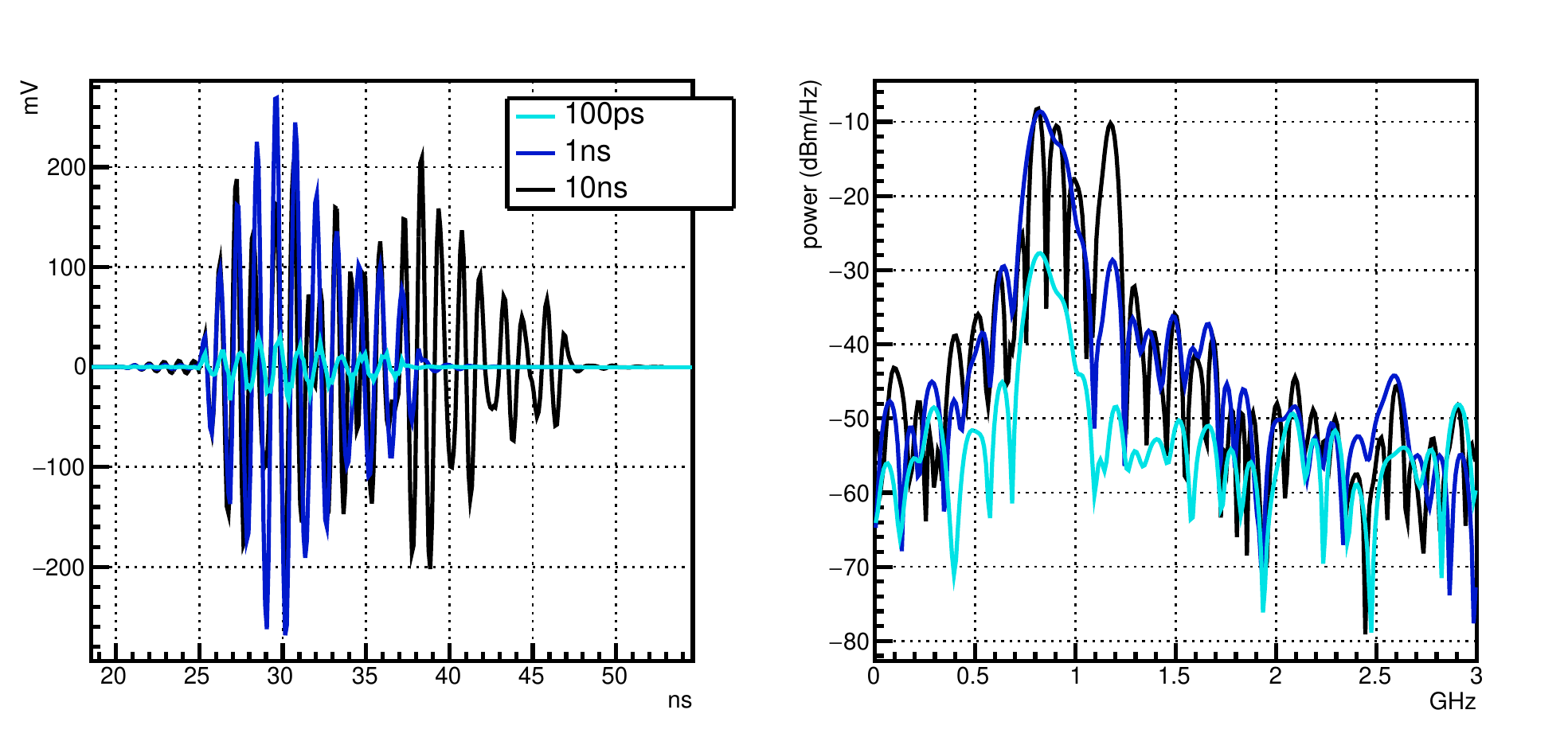}
\par\end{centering}
\caption{Time-domain signals (left) and frequency spectra (right) for various user-defined lifetimes in RadioScatter. The interrogating frequency is 1.15~GHz, and the geometry (Figure~\ref{example_geometry}) is such that we expect chirp-like behavior during shower progression. }
\label{lifetime}
\end{figure}

The plasma lifetime $\tau$ is user-defined at runtime in RadioScatter. Changing $\tau$ changes the phase relationships for re-radiation of the incident signal as a function of time. Using the same TX and RX geometry as in Figure~\ref{example_geometry} and a sounding frequency of 1.15~GHz, the resultant signals for $\tau=$~100~ps, 1~ns, and 10~ns are given in Figure \ref{lifetime}.  The chirp-like frequency shift, an expected function of the geometry of the setup and the progression of the shower, is observed for all lifetimes. The duration of the return signal scales with lifetime, with the spectrum becoming more dominated by the carrier frequency as the lifetime increases. This is fully expected, as the carrier component of the Fourier spectrum is increasingly well-defined with more cycles. That is, the return signal becomes dominated by reflection from a stationary conductor as the plasma lifetime increases. Comparison of empirical results, derived from our testbeam experiment, with these simulation signals will provide experimental bounds on $\tau$.

\subsection{Collisions}\label{col_section}
Collisional effects, which become evident at primary energies $>10^{16}$~eV, and should roughly scale with density, are a further unknown in the model.
The three dominant collision species are electron-electron% (``e-e'')
, electron-ion, and electron-neutral.

We employ Eq.~\ref{collision}, using simple atomic and molecular cross-sections for the $\sigma_s$ terms. %a convenient approximation for the collision frequency, taken from the plasma physics literature\cite{cravens}, 
% \begin{equation}
%54 \frac{n_e}{(e_i/k_B)^{1.5}}
% \end{equation}
% 
% where $e_i$ is the ionization energy of the medium and $k_B$ is Boltzman's constant. 
In general, the dominant collisional species in a plasma is a function of the degree of ionization of the medium. For a dense material such as ice, the number density of neutral, non-ionized molecules exceeds the number density of free charges by several orders of magnitude, so it is likely that the electron-neutral collision rate dominates. But, because the transport and collision rates are not well-known for ice, we calculate the collision frequency using the molecular cross section of water~\cite{water_cross_section}, and, in the absence of experimental data, multiply by a factor of three to conservatively account for all species, including ions and electrons. 
%In typical plasmas, electron-neutral interactions are in fact highly suppressed, % as collisions with neutral molecules rely upon induced dipole moments, 
%but we have conservatively weighted these the same as e-e and e-i collisions.
Our testbeam experiment measures the sum of these three collisional effects.

\subsection{Antenna response}
RadioScatter allows the user to input an antenna gain pattern as a text file with gain, specified separately for TX and RX, as a function of polar and azimuthal angles. If no such antenna pattern is used, the antenna effective height\cite{rice_detector} is set at $\lambda$, essentially making it an idealized antenna with dipole gain at every frequency. This can of course be changed by the user. %The module then includes this gain for the transmitter and receiver, which can be different, with respect to the signal at that point. It is treated as a linear scaling of the amplitude of the received signal. 
Planned for future releases of RadioScatter is an antenna system response that can be convolved with the received signal. This response can be a complex effective height, or a group delay, or an impulse response--i.e., all the variables which characterize the dispersion and amplitude response of an antenna. %We note that, unlike Askaryan detectors, which are sensitive to the peak Cherenkov voltage, summing over frequency, the RadioScatter approach is considerably less sensitive to antenna dispersive effects, e.g.

\section{Upcoming experimental test}
\begin{figure}[H]
\begin{centering}
\includegraphics[width=\textwidth]{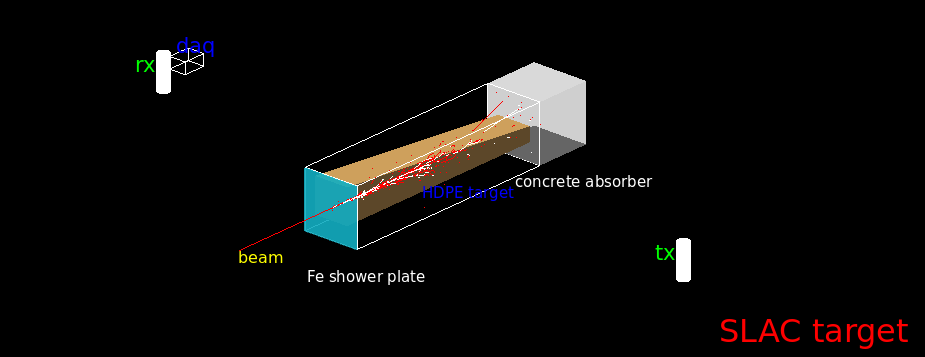}
\par\end{centering}
\caption{GEANT4 representation of the SLAC beam line test, showing a particle shower inside of the HDPE target. The size and type of antennas are not to scale, although the relative distances are approximately accurate for an interrogation frequency of 2~GHz.}
\label{slac}
\end{figure}

The end station test beam (ESTB) facility at the SLAC National Accelerator Laboratory is a user facility which allows researchers to install targets and detectors downstream of a $\cal{O}$(1 Hz) switched electron beam (roughly $10^9$ 10~GeV particles per bunch) from the main linear accelerator. We have proposed using the well-characterized T-510 experiment\cite{t510} target of high-density polyethylene (HDPE) to approximate an in-situ PSP mimicking that of a neutrino/ice interaction. We will then interrogate the PSP within the HDPE target with CW radio, and measure the scattered RF signal. Figure \ref{slac} shows the experimental setup, which was originally designed and optimized for measuring the combined Askaryan and geomagnetic emissions from air showers.
%The unusual shape of the target is due to a design consideration from the experiment for which it was originally designed, SLAC T-510 \cite{t510}, where the combined Askaryan and geomagnetic transmission effect of particle showers was first measured. We will use the same target, as its RF properties have been well characterized by the T-510 collaboration.

This experiment, T-576, is tentatively scheduled for mid-2018. The expected signal for the configuration shown in Figure~\ref{slac} is presented in Figure~\ref{sig}, with separation distances as given in Figure~\ref{example_geometry}.

\part*{Expected Science Reach}
We now consider the radar signals from showers induced by high energy neutrino collisions in ice. In what follows, the transmitting frequency is 450~MHz unless otherwise stated, and, for distant neutrino interactions, the measured attenuation length $L_A=1/\xi$ of ice \cite{barwick} is used in all calculations. Additionally, a plasma lifetime of $\tau_p=$1~ns is used for all calculations. %We note that the 10-kW, 10 MHz SuperDARN transmitter at South Pole would result in signals observable with the surface antennas deployed with ARA stations 1, 2, and 3.

\begin{figure}[H]
\begin{centering}
\includegraphics[width=\textwidth]{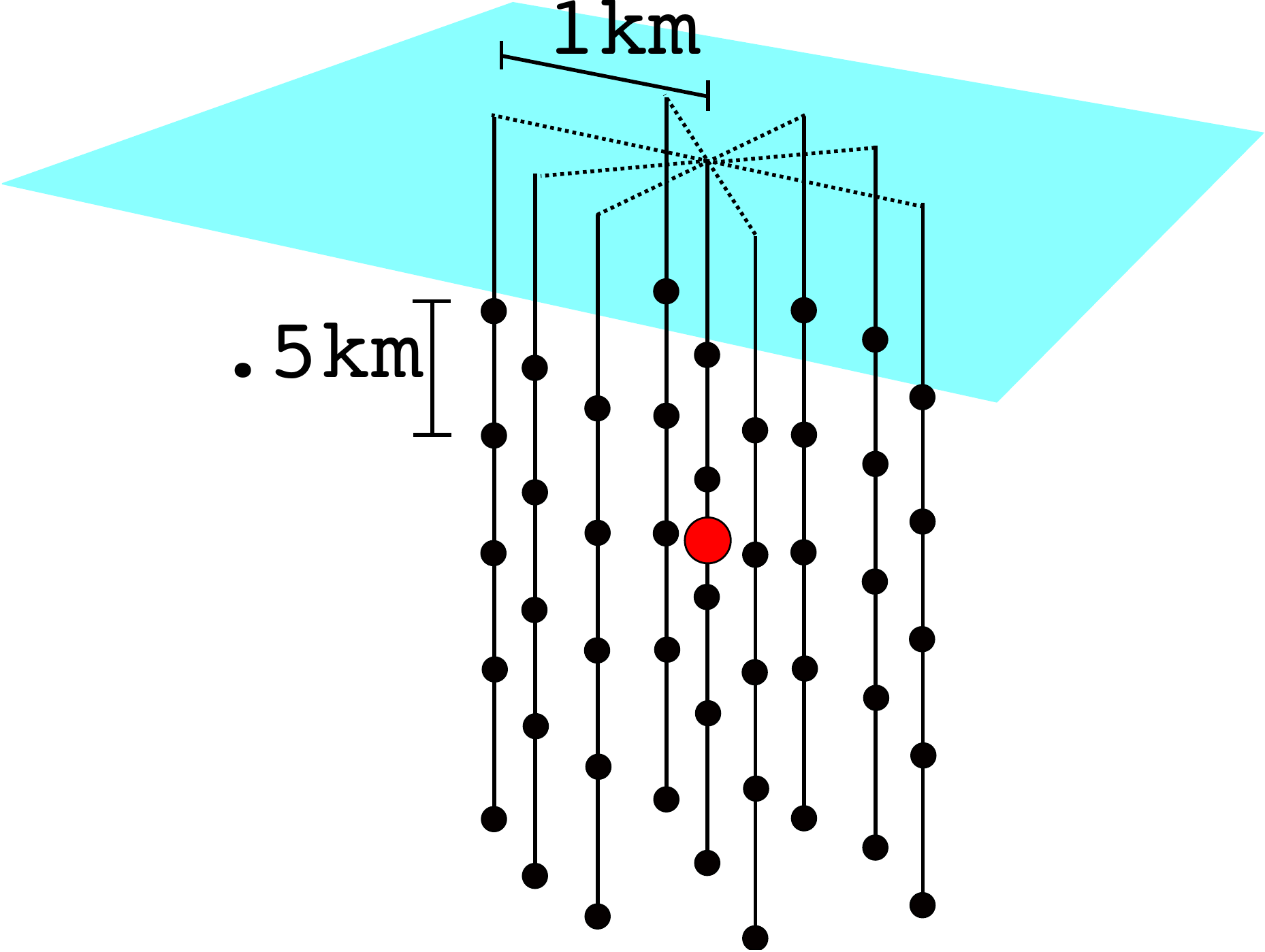}
\par\end{centering}
\caption{An example detector setup for an in-ice radio scatter system. The larger red circle is the transmitter, and the smaller black circles indicate receivers. This station spacing is largely based on measurements of radio attenuation length in ice, to maximise effective volume.}
\label{station_diagram}
\end{figure}

\section{Effective Detector Volume}

The main advantage of the radar technique over current strategies is the ability to scale up the transmitter power, and thus effectively increase the volume of ice in which a candidate neutrino signal may be detected. Since RF/optical experiments detect signals produced by particle showers, as opposed to detecting the shower particles directly, an energy-dependent ``effective volume'' quantifies the amount of sensitive target material accessible to a given detector. %Effective volume is presented as a function of the primary particle energy, since showers of different energies will produce scattered RF of different amplitudes, which can be detected at commensurately different distances. 

\begin{figure}[H]
\begin{centering}
\includegraphics[width=\textwidth]{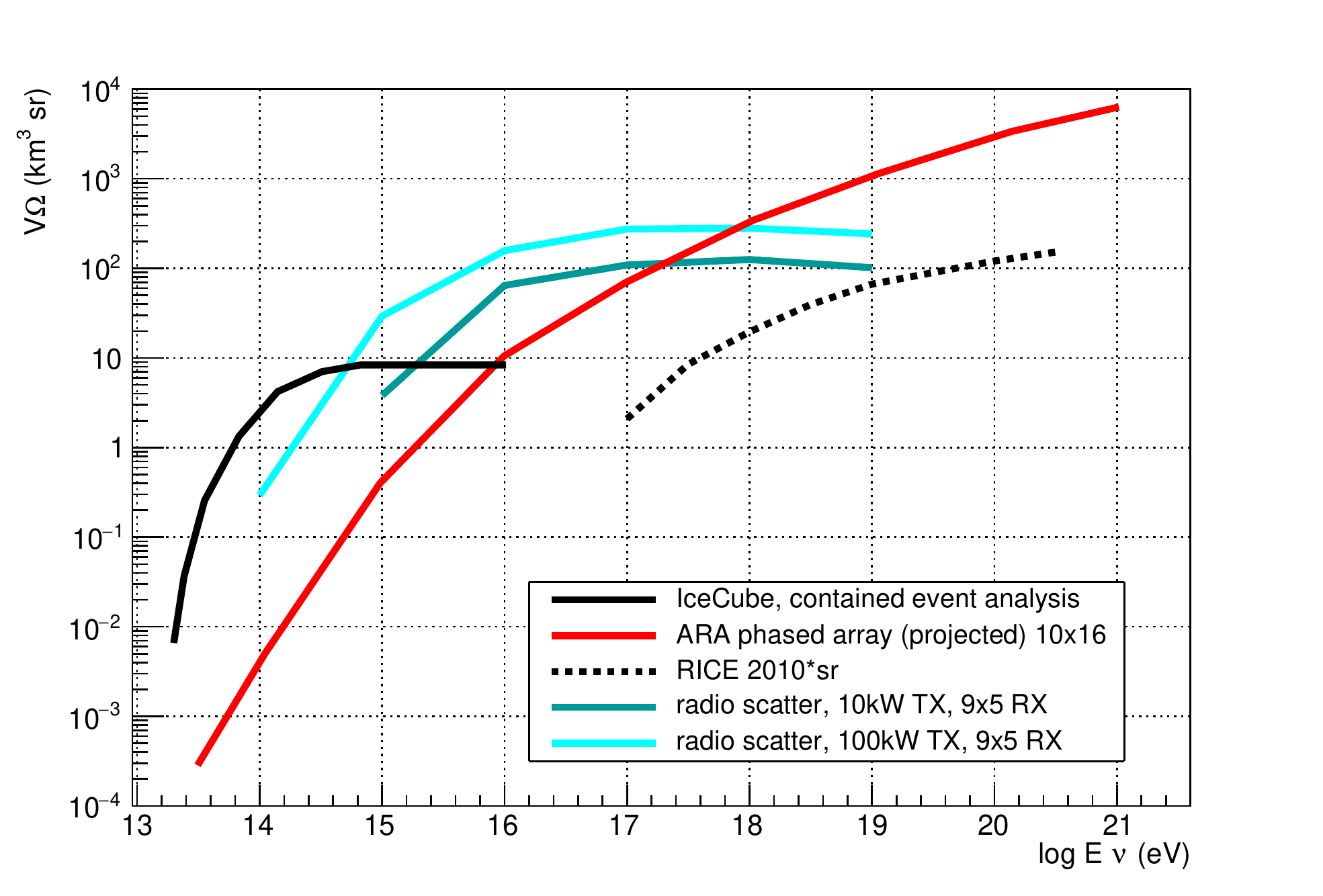}
\par\end{centering}
\caption{Effective volume for a radio scatter experiment, for SNR=3, as a function of primary particle energy, for a 1 TX, 9 station configuration. Each station is a vertical string of 5 antennas. The geometry is shown in Figure~\ref{station_diagram}. Curves correspond to fixed transmitter output power. For comparison, we also show the effective volume for RICE (reproduced from \cite{rice}), IceCube, and the projection for the ARA phased array with a ten station (16 phased antennas per station) configuration(reproduced from \cite{phased_array}).}
\label{thresholds}
\end{figure}

Figure~\ref{thresholds} shows the effective volume of a proposed radio scatter experiment in ice for various values of transmitter output power. The TX-RX configuration for Figure~\ref{thresholds} is shown in Figure~\ref{station_diagram}, and consists of a single transmitter surrounded by 45 receiving antennas, 5 on each of 9 `strings'. To produce this plot, $N(E)$=5000 showers were produced at each decade of energy from $10^{14}$~eV to $10^{19}$~eV and distributed randomly within a $V$=10$\times$10$\times$2.8~km volume, to mimic the ice sheet at the South Pole. The effective volume $V_{eff}(km^3sr)$ at each point in energy $E$ is given by Eq.~\ref{effvol}.
\begin{equation}\label{effvol}
V_{eff}=2\pi V \frac{n(E)}{N(E)}.
\end{equation}
Here, $n(E)$ is the number of events detected at each energy. For simplicity we use a solid angle factor of 2$\pi$ instead of 4$\pi$ to restrict our study to down-going neutrinos (given Earth absorption), and assume a uniform distribution of interaction points within the target volume.
%isotropy of neutrino interaction probabilities in this region, ignoring those that would have passed through the Earth. The angular distribution of arrival directions for simulated showers was also subject to this same restriction.
We set an edge detection threshold of 45~$\mu$V at each receiver, corresponding to a signal-to-noise ratio (SNR) against thermal noise for a 1.2~GHz bandwidth of roughly SNR$\sim$3, and, given the characteristic signature of radar signals, we consider an event to be ``detected'' if any of the antennas trigger at this level. We mention that trigger SNR thresholds of 1:1 have been achieved in experiments designed to detect radar reflections from extensive air showers\cite{tara_limit}. For a 9-station deployment around a single, centrally located 10~kW transmitter, a radio scatter experiment is projected to have greater sensitivity than IceCube above $1$~PeV, and the projected sensitivity of the newly-deployed ARA phased array\cite{phased_array} up to $\sim$0.5~EeV. The increase in effective volume over current strategies is even more pronounced by raising the transmitter power to 100~kW (the typical output power for a terrestrial FM radio station). The radio scatter method is therefore a potential technique for bridging the gap between existing optical and RF detection schemes, essential to establishing the neutrino flux spectrum above 1~PeV\cite{thomas_thesis}.

Not included in the calculation (at the time of this writing) is a full treatment of the bending of rays in the slowly changing index of refraction over the upper $\sim$200~m of the Antarctic ice sheet. This is a geometric effect which will primarily result in re-distribution of signal flux and the presence of some shadow zones at horizontal viewing angles\cite{ara_2_stations}. We have therefore placed our receivers and transmitter in deep ice ($>200$~m deep), where the index of refraction is nearly constant, to mitigate the effect of such ray bending in the simulation.

\section{Geometric acceptance}\label{section:geometric}
The geometric acceptance for the radio scatter technique is perhaps the most compelling rationale for further development of the technique, and is largely responsible for the apparent advantage over Askaryan detectors at $<$~EeV energies. The Askaryan signal exploited by current experiments is forward-beamed, with measurable amplitudes constrained to the Cherenkov angle, corresponding to a restricted geometric aperture\cite{t486}. %By contrast, all of the individual ionization electrons within the PSP cloud are essentially separate dipole radiators polarized along the polarization axis of the TX antenna, for energies $<$0.1~EeV. Above this energy, the collective effects of the dense plasma result in mirror-like scattering, with reflection increasingly localized to the specular reflection angle, as detailed in \cite{krijn_radar_18}.
By contrast, the radar scatter is more isotropic, with measurable returns over a large portion of solid angle for a given shower direction and transmitter location. The reflection is increasingly localized to the specular reflection angle as $\tau$ and  energy increase (e.g. anything that increases the length of the PSP, as detailed in \cite{krijn_radar_18}), but for $<$10~ns lifetimes and $<$~EeV energies, the advantage in geometric acceptance over Askaryan is pronounced.

\begin{figure}[H]
\begin{centering}
\includegraphics[width=\textwidth]{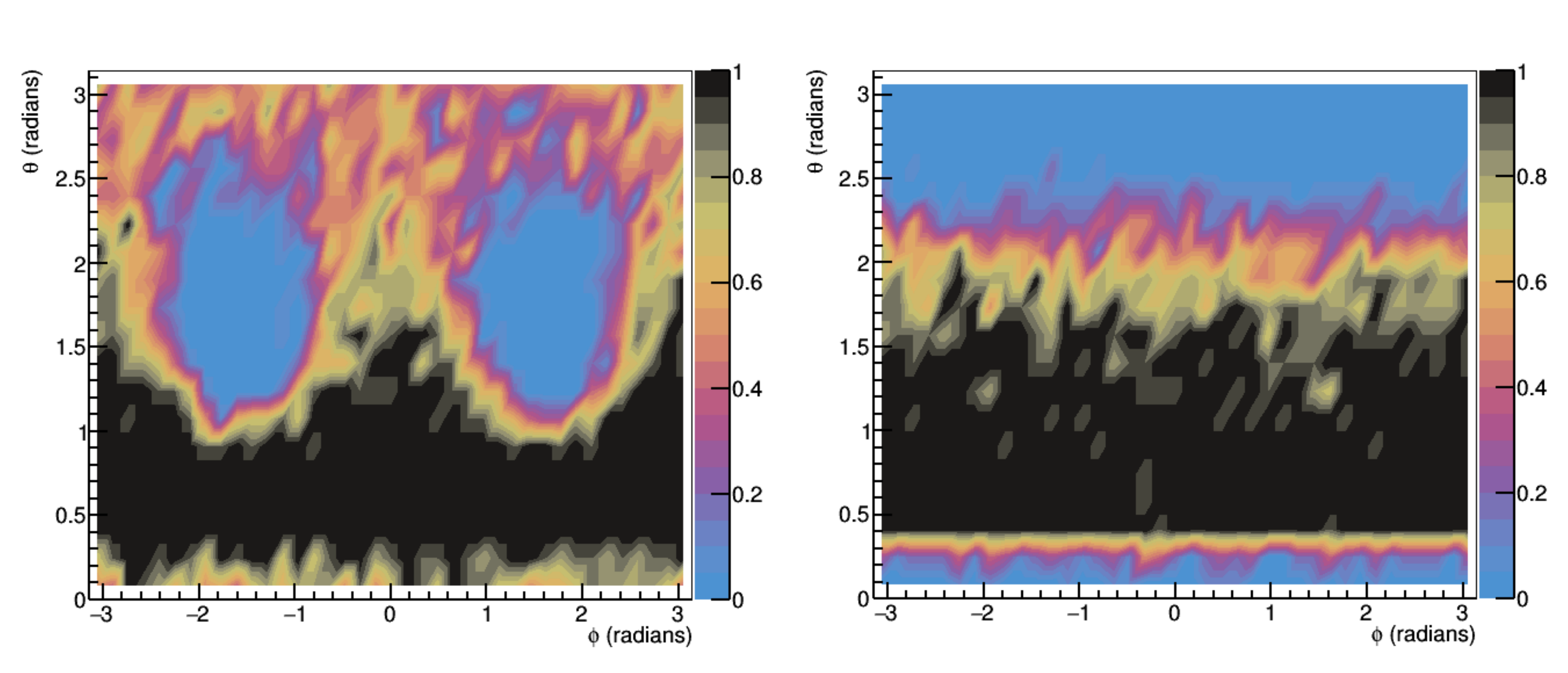}
\par\end{centering}
\caption{Trigger efficiency maps for a requirement of SNR$\geq$10 as a function of angle for a $10^{16}$~eV primary $\nu$ at a radial distance of 1~km from the shower vertex. Left: vertically polarized TX and RX (perpendicular to, and out of the plane of, the shower axis). Right: horizontally polarized TX and RX (parallel to, and in the plane of, the shower axis). Angle and polarization conventions are shown graphically in Figure~\ref{conventions}.}
\label{geometric}
\end{figure} 

Figure~\ref{geometric} shows the trigger efficiency at SNR$\geq$10 for a $10^{16}$~eV primary $\nu$ with $\tau=$1~ns at a fixed radial distance of 100~m from the receiver as a function of spherical coordinates $\phi$ and $\theta$. To produce these maps, the transmitter position is fixed at $\mathbf{r_{TX}}=$(100, 0, 0)$m$, the shower is produced at $\mathbf{r_{s}}=$(0, 0, 0)$m$, with it's momentum direction vector $\mathbf{\hat{p_{s}}}$=(0,0,1). The received signal is calculated, sampling in azimuth and elevation, at a fixed radial distance $r_{RX}=$100$m$ from the vertex. A trigger efficiency (i.e. $n/N$, where $n$ is the number detected and $N$ is the number thrown) is then calculated at each point. These threshold maps are very similar to dipole radiation patterns for vertical and horizontal antennas, respectively. We observe that a high percentage of the solid angle map has high trigger efficiency. %The radiation pattern changes for different geometries, 

%At energies above $10^{17}$~eV, where the `overdense' scattering regime dominates, we expect to see increase in received amplitude at the specular reflection point

%\message{I think it would be useful to calculate the sensitivity assuming a 100 meter or 200 meter deep receiver array - I think 2.5 km is likely too deep to be logistically competitive}
\section{Potential experimental realization}
An  in-ice radio scatter telescope could be co-deployed with a proposed future expansion to the current IceCube experiment, with no additional drilling overhead. The geometry of Figure 13, with 9 holes drilled to a depth of  2.5 km, and each of the 8 perimeter holes laterally displaced 1 km from the center hole is roughly commensurate with that Gen-2 proposed upgrade. The transmitter is deployed in the center hole, along with one detector string, each consisting of 5 antennas separated vertically by 500 m. This value is approximately half of the estimated radio-frequency attenuation length in the upper half of the South Polar ice sheet.

%An in-ice radio scatter telescope could be installed with a relative minimum of apparatus, albeit with high logistical overhead for ice drilling. The geometry of Figure~\ref{station_diagram} would require drilling 9 holes to a depth of 2.5~km, with each of the 8 perimeter holes laterally displaced 1~km from the center hole. The transmitter is deployed in the center hole, along with one detector string, each consisting of 5 antennas separated vertically by 500~m. This value is comparable to the radio-frequency attenuation length in cold polar ice.
% 
 For a 10-100~kW transmitter, an isolated location is most desirable, so as to not interfere with other experiments. A remote Antarctic location, such as Dome C, or a location in Greenland may be candidates for such a deployment. In this paper, we have assumed the well-parametrized ice properties measured at South Pole, which also sites other neutrino detection experiments and therefore offers an opportunity for complementarity. A sufficiently deep transmitter at South Pole should not interfere with other experiments, with RF ``leaking'' out to the air only at angles approximately normal to the surface. Transmission from ice to air will be suppressed at more glancing angles, owing to the Fresnel coefficients.

We have only considered CW here so far in this article, but a detailed study of moduation of the transmitted signal will follow. Modulation of the transmitted signal (standard practice in conventional radar systems) is a further way to lower SNR, increase vertex resolution, and increase sensitivity at lower energies. 
 
 Both the transmitter and the detector strings could be solar-powered during the austral summer, and wind-powered in the austral winter. Data may be relayed from the strings to a central hub via microwave ethernet link, or both power and data may be transferred via trenched cables. 
 
We mention that a preliminary implementation of the method could be performed by deploying a single transmitter and incorporating a new firmware module trigger into the existing ARA experiment at South Pole. The ARA array, though not ideally spaced for a radio scatter experiment, covers a sufficiently large area to be sensitive to the radio scatter method. % with high efficiency. %The effective volume for 3 ARA stations employing the vertically polarized antennas only, with a 10~kW transmitter placed centrally between the 3 ARA stations is shown in Figure~\ref{ara}. The sensitivity of such a setup is comparable to a dedicated 10~kW, 5 station $\times$ 3 antenna configuration. 
% \begin{figure}[H]
%\begin{centering}
%\includegraphics[width=\textwidth]{eff_vol_ara}
%\par\end{centering}
%\caption{Effective volume for a radio scatter experiment, assuming a Signal-to-Noise Ratio threshold of 10:1, as a function of primary particle energy, for a 5 station, 3 antennas per station configuration, and also for a 1~kW transmitter combined with 3 of the existing ARA receiver stations (vertical polarization only). Also shown is $V_{eff}$ for RICE (reproduced from \cite{rice}), IceCube, and the projection for the ARA phased array with a 10 station $\times$ 16 phased array (reproduced from \cite{phased_array}).}
%\label{ara}
%\end{figure}
  Details of an implementation for ARA will be presented in a companion article.

\section{Discussion and outlook}

We have presented a particle-level model for radio/PSP interactions that can be simply incorporated into a GEANT4 simulation via the software module RadioScatter. We have shown that the sum of reflections from individual scatterers results in an appreciable scattered signal amplitude with coherent phase. We have included the effect of plasma screening and collisions, and observe appreciable signal amplitudes for plasma lifetimes as short as $\cal{O}$(100~ps). An in-ice detector with a single 10~kW transmitter has been presented, which has higher calculated sensitivity to neutrinos between 1~PeV and 1~EeV than current optical and Askaryan detectors. This model will be tested in a test-beam experiment at SLAC, planned for spring, 2018. Many of the unknowns in the problem, including the plasma lifetime $\tau$, are direct observables in this experiment. Pending experimental verification, we hope that the radio scatter method can be incorporated into future high energy neutrino detector designs.

\section*{Acknowledgments}
This work is supported by the U.S. National Science Foundation Grant nos. NSF/PHY-0969865 and NSF/MRI-1126353 and a US Department of Energy Office of Science Graduate Student Research (SCGSR) award. The SCGSR program is administered by the Oak Ridge Institute for Science and Education for the DOE under contract number DE‐SC0014664. The authors would like to thank K. de Vries for many fruitful discussions, A. Connolly for invaluable edits, and S. Wissel for ongoing correspondence regarding the work presented here. We would like to thank the reviewers for their thorough and insightful comments. Additionally, SP would like to thank T. Meures, C. Deaconu, and E. Oberla for the initial discussion of this test-beam measurement.

\bibliographystyle{elsarticle-num}

\section*{References}
\bibliography{/home/natas/Documents/physics/tex/bib}{}

%\end{multicols}
\end{document}